\documentclass[pra,twocolumn,showpacs]{revtex4}
\usepackage{epsfig}
\usepackage{amsmath}
\usepackage{graphicx}
\newcommand{\BIG}[2]{\mbox{$\left#1\vbox to #2{}\right. $}}
\newcommand{\half}{\tfrac{1}{2}}

\newcommand{\Li}{ \mbox{Li}}
\newcommand{\Cl}{ \mbox{Cl}}

\newcommand{\sign}{\mbox{sign}}
\DeclareGraphicsExtensions{.png,.jpg}
\begin{document}
 
\title{Recurrence Formulas for Fully Exponentially Correlated Four-Body
Wavefunctions}

\author{Frank E. Harris}
\email{harris@qtp.ufl.edu}
\affiliation{Department of Physics, University of Utah, Salt Lake City,
 Utah 84112}
\affiliation{Departments of Physics and Chemistry, Quantum Theory Project,
 University of Florida, P. O. Box
118435, Gainesville, Florida 32611}
\date{\today}

\begin{abstract}
Formulas are presented for the recursive generation of four-body integrals
 in which the integrand consists of arbitrary integer powers ($\ge -1$)
 of all the interparticle distances $r_{ij}$,
 multiplied by an exponential containing an arbitrary
linear combination of all the $r_{ij}$.   These integrals are generalizations
of those encountered using Hylleraas basis functions, and include all that
are needed to make energy computations on the Li atom and other four-body
 systems with a fully exponentially correlated Slater-type basis
 of arbitrary quantum numbers. The only quantities needed to
start the recursion are the basic four-body 
integral first evaluated by Fromm and
Hill, plus some easily evaluated three-body ``boundary'' integrals.  The
computational labor in constructing integral sets for practical computations
is less than when the integrals are generated using explicit formulas
obtained by differentiating the basic integral with respect to its
parameters.  Computations are facilitated by using a symbolic algebra
program ({\sc maple}) to compute array index pointers and present
syntactically correct {\sc fortran} source code as output; in this way it is
possible to obtain error-free high-speed evaluations with minimal effort.
  The work can be
checked by verifying sum rules the integrals must satisfy.

\end{abstract}

\maketitle

\section{Introduction}

As long ago as 1929, Hylleraas \cite{Hylleraas} presented a computation
of the electronic structure of the He atom showing that a basis of
 explicitly
 correlated wavefunctions provided a far more efficient representation of
that system than was available
from a conventional orbital basis.  What has come to be known as a Hylleraas
atomic basis consists of functions, each of which is
a product of exponentials in the
electron-nuclear distances (often kept identical for all basis members)
to which is appended a product of powers of both the electron-nuclear
and the electron-electron distances.  Hylleraas-basis computations of the
electronic structure and properties of two-electron systems (i.e.\ the
He isoelectronic series) have by now been successfully carried out to
great precision by the inclusion of up to several thousand terms in a
basis-set expansion.  Representative results are those of Yan and Drake
\cite{YanDrake,Drake}.\\ [-6pt]

An alternative to the traditional Hylleraas expansion is the use of
basis functions that have correlation in the exponential, i.e.\ in
which both the electron-nuclear and electron-electron distances appear
exponentially.  This type of basis exhibits (at modest expansion lengths)
an even more efficient representation of two-electron problems than does
the original Hylleraas basis, and has in addition the theoretical advantage
that, because it provides similar descriptions of all the particle
pairs, it is also applicable to so-called \emph{nonadiabatic} systems in
which all three particles have comparable mass.  Extensive computations
of two-electron systems
in these \emph{exponentially correlated} bases have been reported by a
number of investigators; representative of this work is a contribution
by Frolov and Smith \cite{Frolov}.\\[-6pt]

Calculations in the Hylleraas, the exponentially correlated, and other
bases (e.g.\ containing logarithmic terms \cite{Pekeris}) have now been
 carried out, at
least for the neutral He atom, to truly extreme accuracy.  The situation has
been summarized recently by Schwartz \cite{Schwartz}.\\[-6pt]

A related line of endeavor has been to search for wavefunctions which
yield optimum results when restricted to highly compact forms.  Moderate
success in this direction has been obtained using a basis that takes full
cognizance of the asymptotic and other limiting behavior of the
wavefunction \cite{Toennies}; a greater degree of quantitative success
has been achieved for the He isoelectronic series
by careful optimization of four-term exponentially correlated
functions \cite{FEH1,FEH2}.\\ [-6pt]

Part of the reason for the great success with three-body (two-electron)
problems has been that the necessary integrals for both Hylleraas and
exponentially correlated functions are relatively simple, and the
organization of the integral computations has been facilitated by the
existence of recursive procedures \cite{Sack} enabling integrals of the form
\begin{eqnarray}
\Gamma_{n_1,n_2,n_{12}}(\alpha,\beta,\gamma)&=&
\int\! \frac{d{\bf r_1}d{\bf r_2}}{16\pi^2}\,
 r_1^{n_1-1} r_2^{n_2-1} r_{12}^{n_{12}-1}
\nonumber \\ [8pt] &&
\times\; e^{-\alpha r_1-\beta r_2-\gamma r_{12}} \label{e1}
\end{eqnarray}
to be constructed systematically from those of smaller $n_1,n_2,n_{12}$.
Here $r_i$ is the magnitude of ${\bf r}_i$ and $r_{ij} \equiv
|{\bf r}_i-{\bf r}_j|$. \\[-6pt]

The situation becomes drastically different for problems containing
more than three particles.  Historically, integrals for
fully exponentially correlated wavefunctions were regarded as intractable,
while the corresponding integrals for Hylleraas wavefunctions could only
be evaluated by writing the pre-exponential powers of the inter-electron
coordinates as spherical-harmonic expansions \cite{Sack2}.
Thus far, the most accurate studies of a four-body system, the Li atom,
have used the Hylleraas basis set.  A good survey of the current situation
is in a review by King \cite{King}, to which should be
added a recent paper by Puchalski and
Pachucki \cite{Pach2} that reports a more accurate Li computation than
those discussed by King.
\\ [-6pt]

A major advance for four-body systems occurred when
 Fromm and Hill \cite{Hill} presented in 1987
 a closed formula
for the basic exponentially correlated integral
\begin{eqnarray}
I_0(u_1,u_2,u_3,w_1,w_2,w_3) &=&
\int \frac{d{\bf r}_1\, d{\bf r}_2 \,d{\bf r}_3}{64\pi^3} \nonumber \\ [8pt]
&& \hspace{-120pt} \times\;
\frac{e^{-w_1 r_1-w_2r_2-w_3r_3-u_1 r_{23}-u_2 r_{13}-u_3 r_{12}}}
{r_1r_2r_3r_{23}r_{13}r_{12}}\, .
\label{e2}
\end{eqnarray}
However, the
Fromm-Hill formula, though truly a mathematical {\it tour de force}, was
difficult to use, even after simplifications introduced by the present
author \cite{PRA,HFS}, and it was fortunate that in 1991
Remiddi \cite{Remiddi} provided
a much simpler formula for the basic four-body Hylleraas integral (that
of Eq.~(\ref{e2}) with the parameters $u_i$ set to zero). \\[-6pt]  

While the original Fromm-Hill formula could be differentiated with respect to
the $w_i$ and $u_i$ to introduce pre-exponential powers of the $r_i$ and
$r_{ij}$, the lack of $u_i$ dependence in the Remiddi formula made such an
approach unavailable there.  This difficulty was removed when Pachucki,
Puchalski, and Remiddi \cite{Pachucki} published a set of recurrence
relations enabling arbitrary increases to all the pre-exponential powers in
four-body Hylleraas integrals.  While Pachucki {\it et al.}\ indicated that
the extension of their results to the fully exponentially correlated case
would be ``of great interest'', they did not consider that problem in their
work. \\[-6pt]

The present work builds upon a preliminary study by the present author
\cite{FEHrecur} which presented some identities (which could be characterized
as sum rules) connecting four-body exponentially correlated integrals
with contiguous pre-exponential powers. The main result of the present
communication is a family of recurrence formulas which enable construction
of exponentially correlated integrals with arbitrary pre-exponential
powers, starting from the basic integral, Eq.~(\ref{e2}),
 and ``boundary'' integrals
involving fewer than four particles.  It thus consitutes a generalization
of the valuable result of Pachucki {\it et al}. \\ [-6pt]

While the integrals explicitly discussed in this paper [i.e., those represented
by Eq.~(\ref{e3})] involve only the interparticle distances and are therefore
independent of the coordinates needed to describe the overall angular
dependence of a four-body wavefunction, it was pointed out by Fromm and
Hill \cite{Hill} that if spherical-harmonic angular functions are included,
 integration over
their coordinates can be carried out, leaving resultant forms that
can be identified as cases of Eq.~(\ref{e3}).  Details of this reduction
have been addressed in previous work by the present
 author \cite{Angular1,Angular2},
so that in principle the technology to address $P$, $D$, {\dots} states
is complete.  However, the evaluation of the angular contributions to the
kinetic-energy matrix elements is complicated when expressed in terms of the
interparticle coordinates, and there is room for further analysis to
identify straightforward methods for treating these states.

\section{Problem Formulation} 

The integrals that are the subject of this study are of the general form
\begin{eqnarray}
f_{n_1,n_2,n_3,m_1,m_2,m_3}(u_1,u_2,u_3,w_1,w_2,w_3) &=& \nonumber \\ [8pt]
&& \hspace{-194pt}
\int\!\! \frac{d{\bf r}_1\, d{\bf r}_2 \,d{\bf r}_3}{64\pi^3}\,
 r_1^{m_1-1}\,r_2^{m_2-1}\,r_3^{m_3-1}\,
r_{23}^{n_1-1}\,r_{13}^{n_2-1}\,r_{12}^{n_3-1} \nonumber \\ [8pt]
&& \hspace*{-190pt}\times \;
e^{-w_1 r_1-w_2r_2-w_3r_3-u_1 r_{23}-u_2 r_{13}-u_3 r_{12}} \, ,
\label{e3}
\end{eqnarray}
and can be interpreted as describing the interaction of one particle
(Particle 0), at the origin of the coordinate system,
with three others (1,2,3) at the respective positions ${\bf r}_i$.
The integrals $f$ therefore have not only the symmetry corresponding
to renumberings of Particles 1, 2, and 3, but also that resulting
from rewriting Eq.~(\ref{e3}) to place a particle other than Particle 0
at the coordinate origin.  Specifically, the renumbering of 1--3 yields
the identities
\begin{eqnarray}
f_{n_1,n_2,n_3,m_1,m_2,m_3}(u_1,u_2,u_3,w_1,w_2,w_3) &=& \nonumber \\ [8pt]
 f_{n_2,n_1,n_3,m_2,m_1,m_3}(u_2,u_1,u_3,w_2,w_1,w_3)&=& \nonumber \\ [8pt]
 f_{n_3,n_2,n_1,m_3,m_2,m_1}(u_3,u_2,u_1,w_3,w_2,w_1)&=& \nonumber \\ [8pt]
 f_{n_1,n_3,n_2,m_1,m_3,m_2}(u_1,u_3,u_2,w_1,w_3,w_2)&=& \nonumber \\ [8pt]
 f_{n_2,n_3,n_1,m_2,m_3,m_1}(u_2,u_3,u_1,w_2,w_3,w_1)&=& \nonumber \\ [8pt]
 f_{n_3,n_1,n_2,m_3,m_1,m_2}(u_3,u_1,u_2,w_3,w_1,w_2). \label{e4}
\end{eqnarray}
The placement of a particle other than Particle 0
at the coordinate origin yields the additional relations
\begin{eqnarray}
f_{n_1,n_2,n_3,m_1,m_2,m_3}(u_1,u_2,u_3,w_1,w_2,w_3)&=& \nonumber \\ [8pt]
f_{m_1,m_2,n_3,n_1,n_2,m_3}(w_1,w_2,u_3,u_1,u_2,w_3)&=& \nonumber \\ [8pt]
f_{m_1,n_2,m_3,n_1,m_2,n_3}(w_1,u_2,w_3,u_1,w_2,u_3)&=& \nonumber \\ [8pt]
f_{n_1,m_2,m_3,m_1,n_2,n_3}(u_1,w_2,w_3,w_1,u_2,u_3), \label{e5}
\end{eqnarray}
and the complete symmetry of the $f$ is the 24-element group (isomorphic
with that of the 6-$j$ symbol) that is the direct
product of the symmetry operations
 identified in Eqs.~(\ref{e4}) and (\ref{e5}).  Notice
that the parameter set $(u_1,u_2,u_3)$ does not have the same symmetry
properties as $(w_1,w_2,w_3)$; the $u_i$ relate to $r_{jk}$ that form a
triangle, while the $w_i$ relate to $r_i$ that form a star. \\ [-6pt]

An important consequence of the symmetry relations is that it is only
necessary to derive one key recurrence formula, which we choose to be that
which increases the index $n_1$ from those of a reference set.
  Formulas for the advancement of all
 the other indices can then be obtained by an appeal to symmetry. \\[-6pt]

It is convenient, following Pachucki {\it et al}, to
define a \emph{shell} of integrals as those
with a common value of $N\equiv m_1+m_2+m_3+n_1+n_2+n_3$ and refer to $N$
as the \emph{shell index}. We shall find that the key recurrence formula  
relates one integral in the shell of index $N\!\!+\!\!1$
 to a number of integrals in shells of index $N$ or less,
 so a systematic procedure for generating
 integrals in shell $N\!\!+\!\!1$ will involve the prior
generation of all integrals in the shells with indices $\le N$. \\ [-6pt]

Because the number of parameters and indices is rather large, increased
compactness and clarity in the exposition can be achieved by the judicious
use of notational conventions.  We therefore introduce the notion of a
reference index set $n_1,n_2,n_3,m_1,m_2,m_3$ and adopt the convention that
when ambiguity will not thereby result, indices having their reference
values will be omitted.  We also suppress the parameters $u_i$ and $w_i$
whenever possible.  Thus, for example, 
\begin{eqnarray}
 f &\equiv& f_{n_1,n_2,n_3,m_1,m_2,m_3} \, , \\ [8pt]
 f_{n_2+1,m_3-1} &\equiv& f_{n_1,n_2+1,n_3,m_1,m_2,m_3-1} \,, \\ [8pt]
 f_{n_1+1,m_3=1} &\equiv& f_{n_1+1,n_2,n_3,m_1,m_2,1}.
\end{eqnarray}

The recurrence formula we shall derive is most directly formulated
in a notation in which the boundary integrals entering the formula
 are identified as
degenerate cases of the $f$.  Accordingly, using a notation introduced
by Pachucki {\it et al}, we define

\begin{eqnarray}
f_{*,n_2,n_3,m_1,m_2,m_3}&=&
\int\!\! \frac{d{\bf r}_1\, d{\bf r}_2 \,d{\bf r}_3}{64\pi^3}\,
4\pi \delta({\bf r}_{23}) \nonumber \\[8pt] && \hspace{-70pt} \times \,
r_1^{m_1-1}\,r_2^{m_2-1}\,r_3^{m_3-1}\,r_{13}^{n_2-1}\,r_{12}^{n_3-1}
\nonumber \\[8pt] && \hspace{-70pt} \times \,
e^{-w_1 r_1-w_2r_2-w_3r_3-u_2 r_{13}-u_3 r_{12}} . \label{app28x}
\end{eqnarray}
 Note that in Eq.~(\ref{app28x}), the
asterisk indicates the presence of $4\pi \delta(r_{23})$ in place of
$r_{23}^{n_1-1}\exp(-u_1r_{23})$.  Other placements of the asterisk
 correspond
to making this substitution with respect to other $r_{ij}$ or $r_i$.
The notational convention of the preceding paragraph also applies to
these degenerate $f$, so, for example,
\begin{equation}
f_{n_3+1,m_1=*} \equiv f_{n_1,n_2,n_3+1,*,m_2,m_3}.
\end{equation}

\vspace{0pt}

\section{Recurrence Formula}

\vspace{0pt}

We present here the key recurrence formula, deferring its detailed
derivation to Section \ref{S5}.  
This formula, for $f_{n_1+1,n_2,n_3,m_1,m_2,m_3}$, written
in terms of the reference indices $n_1,n_2,n_3,m_1,m_2,m_3$ and therefore
denoted simply $f_{n_1+1}$, takes the deceptively simple form
\begin{equation}
f_{n_1+1} = \frac{C_1 X_1+C_2 X_2 + C_3 X_3}{D}\, . \label{e8}
\end{equation}
The coefficients
 $C_1$, $C_2$, $C_3$, and $D$ are independent of the index values
and are given by
\begin{eqnarray}
C_1&=&u_1(\mu_{23}^2-4 u_2^2 u_3^2), \label{e9} \\ [8pt]
C_2&=& u_2(2 u_3^2\mu_{12}-\mu_{13}\mu_{23}), \label{e10} \\ [8pt]
C_3&=& u_3(2 u_2^2\mu_{13}-\mu_{12}\mu_{23}), \label{e11}
\end{eqnarray}
\vspace{-18pt}
\begin{eqnarray}
D&=&2 u_2 u_3( u_1^2\mu_{23}^2
   +u_2^2\mu_{13}^2 +u_3^2\mu_{12}^2 \nonumber \\ [8pt]
&&-\,\mu_{12}\mu_{13}\mu_{23}-4 u_1^2 u_2^2 u_3^2)\,, \label{e12}
\end{eqnarray}
where the new quantities $\mu_{ij}\rule{0pt}{18pt}$ are

\begin{eqnarray}
 \mu_{12} &=& u_1^2+u_2^2-w_3^2, \label{e13} \\ [8pt]
 \mu_{13} &=& u_1^2+u_3^2-w_2^2, \label{e14} \\ [8pt]
 \mu_{23} &=& u_2^2+u_3^2-w_1^2 \, . \label{e15}
\end{eqnarray}
The numerator quantities $X_i$ in Eq.~(\ref{e8})
depend on the reference index set and on the $f$
from shells of index $\le N$,
thereby imparting the recursive property.  The $X_i$
have the following explicit form, in which the $jk$ sum
is over the two ordered pairs in which $j$ and $k$ are the members
 of (1,2,3) other than $i$, and $\delta_p$ is unity if $p=0$ and zero
otherwise:
\begin{widetext}
\begin{eqnarray}
X_i &=& 2 u_j u_k(n_i+n_j+n_k+1)f - n_j n_k \BIG{[}{10pt}
2u_i f_{n_i+1,n_j-1,n_k-1} - (2n_i+n_j+n_k)f_{n_j-1,n_k-1}\BIG{]}{10pt}
\nonumber \\ [8pt] && \hspace*{-20pt}
+\sum_{jk}\BIG{\{}{14pt}m_j(m_j-1)\BIG{[}{10pt}u_jf_{n_k+1,m_j-2}-n_j f_{n_j-1,n_k+1,m_j-2}\BIG{]}{10pt}
+2n_iu_i \BIG{[}{10pt}u_j f_{n_i-1,n_k+1}-n_j f_{n_i-1,n_j-1,n_k+1}
\BIG{]}{10pt}
\nonumber \\ [6pt] && 
-n_i(n_i-1)\BIG{[}{10pt}u_j f_{n_i-2,n_k+1}-n_j f_{n_i-2,n_j-1,n_k+1}
\BIG{]}{10pt}
-2m_jw_j\BIG{[}{10pt}u_j f_{n_k+1,m_j-1}-n_j f_{n_j-1,n_k+1,m_j-1}\BIG{]}{10pt}
\nonumber \\ [8pt] &&
+n_ju_k \BIG{[}{10pt}2u_i f_{n_i+1,n_j-1}
-(2n_i+n_j+2n_k+1)f_{n_j-1}\BIG{]}{10pt}+n_j\mu_{ik}f_{n_j-1,n_k+1}
\nonumber \\ [8pt] &&
-\delta_{m_j}\BIG{[}{10pt}u_jf_{n_k+1,m_j=*}-n_jf_{n_j-1,n_k+1,m_j=*}
\BIG{]}{10pt}
+\delta_{n_i}\BIG{[}{10pt}u_jf_{n_i=*,n_k+1}-n_jf_{n_i=*,n_j-1,n_k+1}
\BIG{]}{10pt}
\BIG{\}}{14pt}. \label{e16}
\end{eqnarray}
\end{widetext}

\noindent
The last line of Eq.~(\ref{e16}) contains boundary integrals
of the type introduced at Eq.~(\ref{app28x}).  As shown in Appendix B, 
these terms can be written in terms of
the three-body integrals $\Gamma_{n_1,n_2,n_{12}}(\alpha,\beta,\gamma)$
 given in Eq.~(\ref{e1}).
We have
\begin{eqnarray}
f_{*,n_2,n_3,m_1,m_2,m_3} &=& \nonumber \\ [8pt] && \hspace{-90pt}
\Gamma_{m_1,m_2+m_3-1,n_2+n_3-1}(w_1,
w_2\!\!+\!w_3,u_2\!\!+\!u_3),\hspace{20pt} \\ [8pt]
f_{n_1,n_2,n_3,m_1,m_2,*}&=& \nonumber \\ [8pt] && \hspace{-88pt}
\Gamma_{m_1+n_2-1,n_1+m_2-1,n_3}(w_1\!\!+\!u_2,
u_1\!\!+\!w_2,u_3),\hspace{18pt} \label{f_star}
\end{eqnarray}
and further formulas obtainable by simultaneous permutation of the
$u_i$, $w_i$, $m_i$, and $n_i$.\\ [-6pt]

The expressions given above provide a formal route to all
$f$ of shells with $N>0$ from the single basic $N=0$
four-body integral $f_{000000}= I_0$ and various three-body
integrals $\Gamma$.
  To make this paper self-contained, recursive
formulas for the $\Gamma$ are included in Appendix B, and evaluation of
the basic
integral $I_0$ is treated in Appendix C. \\ [-6pt]

The recursive scheme outlined above will fail when the quantity
$D$ or any of its permutational analogs are zero, a condition that
occurs if any of the $u_i$  or $w_i$ vanish.  The methods reported here are
therefore not directly applicable to the Hylleraas basis (in which all
the $u_i$ are zero);  that case is more appropriately handled by the
formulas of Pachucki {\it et al}.

\section{Numerical Evaluation}

It is considerably more complicated than it may at first appear
to make actual calculations based on the recursive process defined in
the preceding section.  Nevertheless, the recursive process turns out to
 be less cumbersome than procedures that depend upon the explicit
evaluation of high-order derivatives of the basic integral presented
as Eq.~(\ref{e2}). \\ [-6pt]

The applications we presently contemplate involve the use of basis sets
that can mimic the $1s^2 2s$ ground-state electronic structure of the Li
atom, and therefore require computations at least as far as the shell
of integrals with $N=8$.  To reach the 
integrals needed from the $N=8$ shell requires the
evaluation of approximately 700 integrals, and it is desirable to carry out
the computations in a way that does not include an unacceptable level
of organizational overhead. \\ [-6pt]

The actual approach we employed was to use {\sc maple} \cite{Maple}
to do the index arithmetic needed to write each specific instance of
Eq.~(\ref{e8}) in a form requiring no index computations, following which
we arranged to have these equations output in a form fully compliant with
{\sc fortan}-95 language specifications and involving no nested loops.
These procedures may seem to be overkill until it is recognized that
index computations may require nearly an order of magnitude more computer
time than the subsequent formation of the recurrence formulas.\\ [-6pt] 

  Grouping the {\sc fortran} formulas into
sets with the same shell index, we were then able to carry out the recursive
computations in a permissible order.  This strategy caused the generation,
through the $N=8$ shell, of nearly 10,000 lines of error-free code.  To
avoid an excessive accumulation of round-off error, all the {\sc fortran}
computations were carried out in quadruple-precision floating point, and
checked for adherence to the sum rules reported in earlier
work \cite{FEHrecur}. The final integral values were generally found consistent
to at least double-precision accuracy.  We note that for the problems for
which the methods of this paper are appropriate, it would not be
cost-prohibitive to carry out the arithmetic operations with even
higher-precision arithmetic.\\ [-6pt]

\section{Derivation of Recurrence Formula \label{S5}}

Following Pachucki {\it et al}.\ \cite{Pachucki}, we introduce a set of
integrals $G$, of definition 
\begin{eqnarray}
G_{n_1,n_2,n_3,m_1,m_2,m_3} &\equiv&
\int  \frac{d{\bf k}_1\, d{\bf k}_2\, d{\bf k}_3}
{8\pi^6} \nonumber \\ [8pt] && \hspace{-112pt}
\times \,(k_1^2+u_1^2)^{-n_1}(k_2^2+u_2^2)^{-n_2}(k_3^2+u_3^2)^{-n_3}
\nonumber \\ [8pt] && \hspace{-112pt} \times \,
(k_{23}^2+w_1^2)^{-m_1}(k_{13}^2+w_2^2)^{-m_2}(k_{12}^2+w_3^2)^{-m_3}.
\nonumber \\ [8pt] &&\hspace{10pt}   \label{e18}
\end{eqnarray}
The relation between $G$ and the integrals $\Gamma$ and $f$,
respectively introduced at Eqs.~(\ref{e1}) and (\ref{e3}),
is discussed in Appendix A; results needed here are
 Eqs.~(\ref{app31})--({\ref{app33}) and their permutational analogs.\\[-6pt]

Continuing the path of Pachucki {\it et al},
we consider the following integral, which can be shown to
vanish by application of Gauss's theorem:
\begin{eqnarray}
I_g&=&\int \frac{d{\bf k}_1\, d{\bf k}_2 \,d{\bf k}_3}{8\pi^6} \nonumber 
\\ [8pt] && \hspace{-20pt} \times \;
\nabla_1 \cdot \BIG{[}{16pt}\frac{{\bf k}_1}{
(k_1^2+u_1^2)(k_2^2+u_2^2)(k_3^2+u_3^2)} \nonumber \\ [8pt] && \hspace{-20pt}
\times\,
 \frac{1}{(k_{23}^2+w_1^2)(k_{13}^2+w_2^2)(k_{12}^2+w_3^2)}\BIG{]}{16pt}.
\hspace{20pt}
\label{e19}
\end{eqnarray}
Carrying out the operations implied by the integrand and identifying the
result in terms of the $G$ (a process that requires the use of identities such
as $2{\bf k}_1\cdot{\bf k}_2=k_1^2+k_2^2-k_{12}^2$),
 we reach
\begin{eqnarray}
I_g&=&2u_1^2G_{211111}+(u_1^2-u_3^2+w_2^2)G_{111121} \nonumber \\ [8pt]
&&\hspace{-20pt}+\,(u_1^2-u_2^2+w_3^2)G_{111112}-G_{111111}-G_{011112}
 \nonumber \\ [8pt]
&&\hspace{-20pt}-\,G_{011121}
+G_{101112}+G_{110121}=0. \label{eqG}
\end{eqnarray}
At this point it is convenient to
modify Eq.~(\ref{eqG}) to a symmetry-equivalent equation
 by interchanging $u_2 \leftrightarrow w_2$,
$u_3 \leftrightarrow w_3$, $n_2 \leftrightarrow m_2$,
 $n_3 \leftrightarrow m_3$, thereby obtaining
\begin{eqnarray} && \hspace*{-18pt}
2u_1^2G_{211111}+\mu_{12}G_{121111}+\mu_{13}G_{112111}-G_{111111}
 \nonumber \\ [8pt] && \hspace*{-18pt}
-\,G_{012111}-G_{011121}+G_{112101}+G_{121110} =0. \hspace{10pt}
\label{e19a}
\end{eqnarray}
We now replace the $G$ by their equivalents in terms of $f$, using 
formulas from Appendix A.  After
multiplying through by the factor needed to clear all variables from
the denominators, Eq.~(\ref{e19a}) becomes
\begin{eqnarray}
2u_1u_2u_3f_{100000}+\mu_{12}u_3f_{010000} \nonumber \\ [8pt]
+\,\mu_{13}u_2f_{001000} &=&\hat{X}_1, \label{e20}
\end{eqnarray}
where
\begin{eqnarray}
\hat{X}_1&=&2u_2u_3f_{000000}+ u_3(f_{*10000}-f_{01000*})\hspace{20pt}
 \nonumber\\[8pt]
&& +\, u_2(f_{*01000}-f_{00100*}).
 \label{e23}
\end{eqnarray}

Our next step is to apply to both sides of Eq.~(\ref{e20})
  the differentiation operator
\begin{displaymath}
{\cal D} = \prod_{i=1}^3 \left(-\frac{\partial}{\partial u_i}\right)^{n_i}
\left(-\frac{\partial}{\partial w_i}\right)^{m_i}\,,
\end{displaymath}
after which we define the reference index values to be
$(n_1,n_2,n_3,m_1,m_2,m_3)$.
Looking at Eq.~(\ref{e3}), we see that differentiation of $f$ with
respect to $u_i$ (or $w_i$) will cause its index $n_i$ (or $m_i$)
to be increased by unity.  Therefore, the left hand side of the resulting
equation will contain one term in which the differentiations are all
applied to the function $f_{100000}$; this term will have the
same coefficient as the $f_{100000}$ term of Eq.~(\ref{e20})
and, in terms of the reference index values, $f_{100000}$ becomes
 $f_{n_1+1}$.  Similar observations apply to $f_{010000}$ and $f_{001000}$.
There will also be additional terms that result when one or more of the
left-hand-side differentiations are applied to the coefficients $2u_1u_2u_3$, 
$\mu_{12}u_3$, or $\mu_{13}u_2$.  We transpose these terms to the
right hand side and combine them
 with the result of differentiating $\hat{X}_1$. \\ [-6pt]

When $\cal D$ is applied to $\hat{X}_1$, we encounter differentiations
of quantities such as $f_{*10000}$.  Keeping in mind that 
$f_{*10000}$ does not depend upon $u_1$ but depends exponentially on
the other $u_i$ and $w_i$, we see that ${\cal D}f_{*10000}$ will vanish
unless $n_1=0$, and nonzero values of the other $n_i$ and $m_i$ will
result in incrementation of the non-asterisked indices.  This lack of
$n_1$ dependence
leads to the introduction of a factor $\delta_{n_1}$ in the
differentiation.  Corresponding observations apply to the other
terms containing asterisks. \\ [-6pt]

Based on the analysis of the preceding two paragraphs,
the application of $\cal D$ to Eq.~(\ref{e20}) can be seen to yield
the first of the three equations shown below.  The second and third
of these equations follow by permutation of the indices in the first
equation.
\begin{eqnarray}
2u_1u_2u_3f_{n_1+1}+\mu_{12}u_3f_{n_2+1}+\mu_{13}u_2f_{n_3+1}
&=& X_1, \nonumber \\ [8pt]
\mu_{12}u_3f_{n_1+1}+2u_1u_2u_3f_{n_2+1}+\mu_{23}u_1f_{n_3+1}
&=& X_2, \nonumber \\ [8pt]
\mu_{13}u_2f_{n_1+1}+\mu_{23}u_1f_{n_2+1}+2u_1u_2u_3f_{n_3+1}
&=& X_3. \nonumber \\ [8pt]
\hspace*{10pt} \label{e22}
\end{eqnarray}
The $X_i$ have the values given in Eq.~(\ref{e16}).\\ [-6pt]

Finally, we solve the equation set, Eq.~(\ref{e22}).  
Applying Cramer's Rule, we get for $f_{n_1+1}$:
\begin{equation}
f_{n_1+1} = \frac{1}{\hat{D}}\left| \begin{array}{ccc}
X_1 & \mu_{12}u_3 & \mu_{13}u_2 \\ [6pt] X_2 & 2u_1u_2u_3 & \mu_{23}u_1 
\\ [6pt] X_3 & \mu_{23}u_1 & 2u_1u_2u_3
 \end{array} \right|, \label{e26}
\end{equation}
with
\begin{equation}
\hat{D} = \left| \begin{array}{ccc}
2u_1u_2u_3 & \mu_{12}u_3 & \mu_{13}u_2 \\ [6pt]
 \mu_{12}u_3 & 2u_1u_2u_3 & \mu_{23}u_1 
\\ [6pt] \mu_{13}u_2 & \mu_{23}u_1 & 2u_1u_2u_3
\end{array}
\right| \label{e27}
\end{equation}
Expanding the determinants and dividing 
 the numerator and denominator of Eq.~(\ref{e26}) by $-u_1$, we obtain
the expression for $f_{n_1+1}$ shown in Eq.~(\ref{e8}),
with $C_1$, $C_2$, $C_3$, and $D$ as given in
 Eqs.~(\ref{e9})--(\ref{e12}).  We need not exhibit solutions for
$f_{n_2+1}$ or $f_{n_3+1}$ because they can be reached by permutation
of the indices in the expression for $f_{n_1+1}$.

\begin{acknowledgments}

This work was supported by the U.S. National Science Foundation,
Grant PHY-0601758.

\end{acknowledgments}

\appendix 

\section{Fourier Representation Formulas}

The formulas in section \ref{S5} have forms that depend crucially on
the Fourier-representation forms of four-body integrals of the generic
type
\begin{eqnarray}
L&=& \int \frac{d{\bf r}_1\, d{\bf r}_2\, d{\bf r}_3}{64\pi^3}
\ \nonumber \\ [8pt]
&& \hspace{-20pt}\times \,
 h_{23}(r_{23})h_{13}(r_{13})
h_{12}(r_{12})h_1(r_1)h_2(r_2)h_3(r_3) \nonumber \\ [8pt]
&=&\int \frac{d{\bf r}_1\, d{\bf r}_2\, d{\bf r}_3}{64\pi^3}
 \int \frac{d{\bf q}_1\,d{\bf q}_2\,d{\bf q}_3\,
 d{\bf q}_4\,d{\bf q}_5\,d{\bf q}_6}{(2\pi)^{18}} \nonumber \\ [8pt]
&& \hspace{-20pt}
\times\, h^T_{23}(q_1)h^T_{13}(q_2)h^T_{12}(q_3)h^T_1(q_4)h^T_2(q_5)
h^T_3(q_6) \nonumber \\ [8pt]
&& \hspace{-20pt} \times 
 \exp\BIG{(}{12pt}i\BIG{[}{10pt}{\bf q}_1\cdot({\bf r}_2-{\bf r}_3)
+{\bf q}_2\cdot({\bf r}_3-{\bf r}_1) \nonumber \\ [8pt]
&&\hspace{-20pt} +\,{\bf q}_3\cdot({\bf r}_1-{\bf r}_2)
-{\bf q}_4\cdot{\bf r}_1-{\bf q}_5\cdot{\bf r}_2-{\bf q}_6\cdot{\bf r}_3
\BIG{]}{10pt}\BIG{)}{12pt}.\nonumber \\ [8pt]
\hspace*{1pt} \label{app23}
\end{eqnarray}
Here $h(r)$ is a direct-space function  and
$h^T(q)$ is its Fourier transform.  The transform pairs needed here are
\begin{eqnarray}
h(r)=\frac{e^{-tr}}{r},&\hspace{16pt}&h^T(q)=\frac{4\pi}{q^2+t^2},
\label{app24} \\ [8pt]
h(r)= \delta({\bf r}),&&h^T(q)=1. \label{app25}
\end{eqnarray}
Performing now the ${\bf r}_i$ integrations, which are all of the
generic type
\begin{equation}
\int e^{i{\bf q}\cdot{\bf r}}d{\bf r}=(2\pi)^3\delta({\bf q}), \label{app26}
\end{equation}
and then evaluating the integrals over ${\bf q}_4$, ${\bf q}_5$, and
${\bf q}_6$, we find
\begin{eqnarray}
L&=&\int \frac{d{\bf q}_1\,d{\bf q}_2\,d{\bf q}_3}{2^{15}\pi^{12}}
h^T_{23}(q_1)h^T_{13}(q_2)h^T_{12}(q_3) \nonumber \\ [8pt] &&
\times\,h^T_1(q_{23})h^T_2(q_{13})h^T_3(q_{12}), \label{app27}
\end{eqnarray}
where $q_{ij}=|{\bf q}_i-{\bf q}_j|$. \\[-6pt]

We now insert into $L$ as given by Eq.~(\ref{app27}), factors $h^T$ of
the form in Eq.~(\ref{app24}), with the result that $L$ becomes equal
to the integral $G_{111111}$ as defined in Eq.~(\ref{e18}).
 In addition, we can insert
the corresponding functions $h$ into
the direct-space form in Eq.~(\ref{app23}), thereby also
identifying $L$ as 
$f_{000000}$, defined in Eq.~(\ref{e3}).  Equating these forms for $L$,
we reach
\begin{equation}
 G_{111111} = f_{000000}. \label{app28}
\end{equation}

Next, we consider the result when we evaluate $L$ taking $h_{23}$ and
$h^T_{23}$ of the form in Eq.~(\ref{app25}), with the other $h$ and
$h^T$ continuing as instances of Eq.~(\ref{app24}).  We then have,
from Eq.~(\ref{app27}), $L=G_{011111}/4\pi$.
Alternatively,  the direct-space
formula for this $L$ can be identified as $f_{*00000}/4\pi$, where
the asterisk-containing $f$ is the
degenerate form introduced at Eq.~(\ref{app28x}).
Equating these alternate forms for $L$, we have
\begin{equation}
G_{011111} = f_{*00000}\,. \label{app29}
\end{equation}

Similar operations can be carried out if $L$ is evaluated taking
 Eq.~(\ref{app25}) for $h_3$ and $h^T_3$, with Eq.~(\ref{app24}) for the
other $h$ and $h^T$.  The result is
\begin{equation}
G_{111110} = f_{00000*}\,. \label{app30}
\end{equation}
Now,
differentiating both sides of Eqs.~(\ref{app29}) and (\ref{app30}) with
respect to $u_2$, we obtain the following results needed in the main text:
\begin{eqnarray}
G_{021111} = \frac{f_{*10000}}{2u_2}, \label{app31} \\ [8pt]
G_{121110} = \frac{f_{01000*}}{2u_2}. \label{app32}
\end{eqnarray}
Finally, we need the result of differentiating Eq.~(\ref{app28}) with
respect to $u_1$:
\begin{equation}
G_{211111} = \frac{f_{100000}}{2u_1}. \label{app33} 
\end{equation}
Results analogous to those in Eqs.~(\ref{app31})--(\ref{app33}) can
be obtained by simultaneous permutation of the first and second
groups of three indices in $f$ and $G$ and the indices of $u$ and $w$.

\section{Three-Body Integrals}

In order to carry out the recursive process defined by Eq.~(\ref{e16}),
we will need to evaluate integrals of the form introduced
in Eq.~(\ref{app28x}).   Carrying out the ${\bf r}_3$ integrations, two
such integrals reduce to the three-body integrals
\begin{eqnarray}
f_{*,n_2,n_3,m_1,m_2,m_3}&=&\int \frac{d{\bf r}_1\,d{\bf r}_2}{16\pi^2}
r_1^{m_1-1}r_2^{m_2+m_3-2} \nonumber \\ [8pt] && \hspace{-80pt}
\times\,r_{12}^{n_2+n_3-2} e^{-w_1r_1-(w_2+w_3)r_2-(u_2+u_3)r_{12}}, \\ [8pt]
f_{n1,n2,n3,m1,m2,*}&=&\!\!\int \frac{d{\bf r}_1\,d{\bf r}_2}{16\pi^2}
r_1^{m_1+n_2-2}r_2^{n_1+m_2-2} \nonumber \\ [8pt] && \hspace{-80pt}
\times\,r_{12}^{n_3-1} e^{-(w_1+u_2)r_1-(u_1+w_2)r_2-u_3r_{12}}.
\end{eqnarray}
These integrals can be respectively identified as
\begin{eqnarray}
&&\Gamma_{m_1,m_2+m_3-1,n_2+n_3-1}(w_1,
w_2\!\!+\!w_3,u_2\!\!+\!u_3),\hspace{20pt} \nonumber \\ [8pt]
&&\Gamma_{m_1+n_2-1,n_1+m_2-1,n_3}(w_1\!\!+\!u_2,
u_1\!\!+\!w_2,u_3),\hspace{18pt} \nonumber
\end{eqnarray}
as shown in Eq.~(\ref{f_star}) of the main text. \\ [-6pt]

The asterisked $f$ needed for the present work are equivalent to
$\Gamma$ in which no more than one of the indices is negative
(with the only negative value $-1$).  The recursive methods most often used for
evaluating $\Gamma$ do not directly permit advancement of an index from $-1$;
we also note that $\Gamma$ is invariant with respect to simultaneous
permutation of its indices and arguments.  We may therefore
identify the $\Gamma$ needed here as falling into two cases:
(1) $\Gamma_{n_1,n_2,n_{12}}$
with all indices non-negative, and (2) $\Gamma_{-1,n_2,n_{12}}$ with
$n_2$ and $n_{12}$ non-negative.\\ [-6pt]

For the first case, the recursive process can start from $\Gamma_{000}$,
which by direct integration is found to have the value
\begin{equation}
\Gamma_{000}(\alpha,\beta,\gamma)=\frac{1}{(\alpha+\beta)(\alpha+\gamma)
(\beta+\gamma)},
\end{equation}
which we rewrite
\begin{eqnarray}
\Gamma_{000}(\alpha,\beta,\gamma)&=&\frac{B_{000}(\alpha,\beta,\gamma)}
{\alpha+\beta},\\ [8pt]
B_{000}(\alpha,\beta,\gamma)&=&\frac{A_{000}(\alpha,\beta,\gamma)}
{\alpha+\gamma}\\ [8pt]
A_{000}(\alpha,\beta,\gamma)&=&\frac{1}{\beta+\gamma}.
\end{eqnarray}
We now introduce
\begin{equation}
{\cal D}_{n_1n_2n_{12}} = \left(-\frac{\partial}{\partial\alpha}\right)^{n_1}
\left(-\frac{\partial}{\partial\beta}\right)^{n_2}
\left(-\frac{\partial}{\partial\gamma}\right)^{n_{12}},
\end{equation}
and apply the recursive procedure of Sack, Roothaan, and Kolos \cite{Sack},
leading to the following formulas:
\begin{eqnarray}
\Gamma_{n_1n_2n_{12}}\hspace{-6pt}&=&\hspace{-1.2pt}{\cal D}_{n_1n_2n_{12}}\Gamma_{000}=
\frac{1}{\alpha+\beta}\BIG{[}{14pt}n_1\Gamma_{n_1-1,n_2,n_3}
 \nonumber \\ [8pt]
&&\hspace{-00pt}
+\,n_2\Gamma_{n_1,n_2-1,n_3}+B_{n_1n_2n_{12}}\BIG{]}{14pt}, \\ [8pt]
B_{n_1n_2n_{12}}\hspace{-6pt}&=&\hspace{-1.2pt}{\cal D}_{n_1n_2n_{12}}B_{000}=
\frac{1}{\alpha+\gamma}\BIG{[}{14pt}n_1B_{n_1-1,n_2,n_3}
\nonumber \\ [8pt]
&&\hspace{-00pt}
+\,n_{12}B_{n_1,n_2,n_{12}-1}
+A_{n_1,n_2,n_3}\BIG{]}{14pt}, \\ [8pt]
A_{n_1,n_2,n_3}\hspace{-6pt}&=&\hspace{-1.2pt}{\cal D}_{n_1n_2n_{12}}A_{000}
\nonumber \\ [8pt]&=
&\frac{\delta_{n_1}(n_2+n_{12})!}{(\beta+\gamma)^{n_1+n_{12}+1}}\, .
\end{eqnarray}
It is a computationally stable procedure to evaluate first array $A$, then
$B$, and finally $\Gamma$.\\

For the second case, namely the integrals $\Gamma_{-1,n_2,n_{12}}$,
 a starting formula, again by direct integration, is
\begin{equation}
\Gamma_{-1,0,0} (\alpha,\beta,\gamma)= \frac
 {\ln(\alpha+\beta)-\ln(\alpha+\gamma)}{\beta^2-\gamma^2}.
\end{equation}
If $\beta-\gamma$ is not too small, one can proceed by a variant of the
procedure of Sack {\it et al}.  Writing
\begin{eqnarray}
\Gamma_{-1,0,0}(\alpha,\beta,\gamma)&=&\frac{G_{00}(\alpha,\beta,\gamma)}
{\beta+\gamma}, \\ [8pt]
G_{00}(\alpha,\beta,\gamma)&=& \frac{K_{00}(\alpha,\beta,\gamma)}
{\beta-\gamma}, \\ [8pt]
K_{00}(\alpha,\beta,\gamma)&=&\ln(\alpha+\beta)-\ln(\alpha+\gamma) \, ,
\hspace{26pt}
\end{eqnarray}
the recurrence formulas become
\begin{eqnarray}
\Gamma_{-1,n_2,n_{12}}&=&\frac{1}{\beta+\gamma}\BIG{[}{14pt}
n_2\Gamma_{-1,n_2-1,n_{12}} \nonumber \\ [8pt]
&&\hspace{-20pt}+\,n_{12}\Gamma_{-1,n_2,n_{12}-1}+G_{n_2n_{12}}\BIG{]}{14pt},
\hspace{20pt} \label{B15} \\ [8pt]
G_{n_2n_{12}}&=&\frac{1}{\beta-\gamma}\BIG{[}{14pt}
n_2G_{n_2-1,n_{12}} \nonumber \\ [8pt]
&&\hspace{-20pt}-\,n_{12}G_{n_2,n_{12}-1} + K_{n_2n_{12}}\BIG{]}{14pt},\\[8pt]
K_{n_2n_{12}}&=&\delta_{n_2}\delta_{n_{12}}K_{00} \nonumber\\[8pt]
&&\hspace{-20pt}
-\,\frac{\delta_{n_{12}}(1-\delta_{n_2})(n_2-1)!}{(\alpha+\beta)^{n_2}}
\nonumber \\ [8pt] &&\hspace{-20pt}
+\,\frac{\delta_{n_2}
(1-\delta_{n_{12}})(n_{12}-1)!}{(\alpha+\gamma)^{n_{12}}}
 \, .
\end{eqnarray}

For $\beta-\gamma$ small, it is more advisable to introduce
 $x=\half(\beta+\gamma)$, $y=\half(\beta-\gamma)$, to write
\begin{equation}
G_{0,0}=\frac{\ln(\alpha+x+y)-\ln(\alpha+x-y)}{2y},
\end{equation}
and to expand in powers of $y$. The result is
\begin{equation}
G_{0,0}=\sum_{k=0}^\infty\;
\frac{y^{2k}}{(2k+1)(\alpha+x)^{2k+1}}\, . \label{B19}
\end{equation}
Differentiation of Eq.~(\ref{B19}) leads to the expansion

\rule{0pt}{20pt}

\hspace*{20pt}

\vspace*{-24pt}
\begin{eqnarray}
G_{n_2n_{12}}\!\!&=& \!\!n_2!\,n_{12}!\BIG{[}{14pt}\frac{1}{n_2+n_{12}+1}
+\,\frac{n_{12}-n_2}{n_2+n_{12}+2}\:y \nonumber \\ [8pt]
&& \hspace{-40pt} +\, \frac{(n_2-n_{12})^2+n_2+n_{12}+2}
{2(n_2+n_{12}+3)}\;y^2 + \dots \BIG{]}{14pt},\hspace{24pt}
\end{eqnarray}
which can then be inserted into Eq.~(\ref{B15}).

\section{Basic Integral $I_0$}

The integral $I_0$, defined in Eq.~(\ref{e2}), is needed to start
the recursive process.  As discussed in \cite{Hill,PRA,HFS},
the evaluation depends upon whether the quantity $\sigma$ is real, where
\begin{eqnarray}
\sigma^2&=&w_1^2w_2^2w_3^2+w_1^2u_2^2u_3^2+w_2^2u_1^2u_3^2+w_3^2u_1^2u_2^2
\nonumber \\ [4pt]
&&\hspace{-20pt}+\,w_1^2u_1^2(w_1^2+u_1^2-w_2^2-u_2^2-w_3^2-u_3^2)
\nonumber \\ [4pt]
&&\hspace{-20pt}+\,w_2^2u_2^2(w_2^2+u_2^2-w_1^2-u_1^2-w_3^2-u_3^2)
\nonumber \\ [4pt]
&&\hspace{-20pt}+\,w_3^2u_3^2(w_3^2+u_3^2-w_1^2-u_1^2-w_2^2-u_2^2).\hspace{20pt}
\end{eqnarray}
For real $\sigma$, $I_0$ is given by
\begin{equation}
I_0=\frac{1}{4\sigma}\BIG{[}{14pt}\!-\!2\sum_{i=1}^3 v\!\left(\frac{\Gamma_i}\sigma
\right)+\sum_{i,j=0}^3 v\!\left(\frac{\gamma_j^{(i)}}{\sigma}\right)
+\frac{\pi^2}{2}\BIG{]}{14pt}, \label{C2}
\end{equation}
where
\begin{eqnarray}
v(z)&=&\sign(z)\BIG{[}{14pt}-\frac{1}{4}\ln^2\left|\frac{1-|z|}{1+|z|}\right|
-\frac{\pi^2}{12} \nonumber \\ [8pt]
&& \hspace{-20pt}+\,\Li_2\left(\frac{1-|z|}{2}\right)+\frac{1}{2}\ln^2\left(
\frac{1+|z|}{2}\right)\BIG{]}{14pt},\hspace{12pt} \label{C3}
\end{eqnarray}
and $\Li_2(z)$ is the dilogarithm (see Formula 27.7.1 of \cite{AS};
also Lewin \cite{Lewin}).  Both the logarithm
and $\Li_2$ are multiple-valued, but, contrary to the original formulation
that required branch tracking
\cite{Hill}, Eq.~(\ref{C2}) can be evaluated straightforwardly with all
functions assigned their principal values.\\

For imaginary $\sigma$, which occurs for physically relevant parameter
values, $I_0$ is obtained from
\begin{eqnarray}
I_0&=&\frac{1}{4|\sigma|}\BIG{[}{14pt}-2\sum_{i=1}^3
 \Cl_2\BIG{(}{10pt}\pi-2\tan^{-1}(\Gamma_i/|\sigma|)\BIG{)}{10pt} \nonumber \\ [8pt]
&&+\sum_{i,j=0}^3\Cl_2\BIG{(}{10pt}\pi-2\tan^{-1}(\gamma_j^{(i)}/|\sigma|)
\BIG{)}{10pt}\BIG{]}{14pt}. \label{C4}
\end{eqnarray}
Here $\Cl_2(\theta)$ is the Clausen function (\cite{AS}, Formula 27.8.1).
Note that because $\Cl_2$ is periodic with period $2\pi$, the presence of
the arctangent does not cause multiple-valuedness in Eq.~(\ref{C4}).\\

The quantities $\Gamma_i$ $(i=1,2,3)$  appearing in
 Eqs.~(\ref{C2}) and (\ref{C4}) are defined as follows:
\begin{widetext}
\begin{equation}
\Gamma_i=\frac{[w_i^2+(u_j+u_k)(w_j+w_k)][u_i^2+(u_j+w_k)(w_j+u_k)]}
{u_j+u_k+w_j+w_k}-(u_j+u_k+w_j+w_k)(u_jw_j+u_kw_k)
\end{equation}
\end{widetext}
The four $\gamma^{(0)}_j$ are
\begin{eqnarray}
\gamma_0^{(0)}&=& 2u_1u_2u_3+u_1\mu_{23}+u_2\mu_{13}+u_3\mu_{12},
\nonumber \\[4pt]
\gamma_1^{(0)}&=& -2u_1u_2u_3-u_1\mu_{23}+u_2\mu_{13}+u_3\mu_{12},
\nonumber \\[4pt]
\gamma_2^{(0)}&=& -2u_1u_2u_3+u_1\mu_{23}-u_2\mu_{13}+u_3\mu_{12},
\nonumber \\[4pt]
\gamma_3^{(0)}&=& -2u_1u_2u_3+u_1\mu_{23}+u_2\mu_{13}-u_3\mu_{12},
\nonumber \\
\hspace*{10pt}
\end{eqnarray}
where $\mu_{ij}$ are as defined in Eqs.~(\ref{e13})--(\ref{e15}).
The $\gamma_j^{(i)}$ with $i\ne 0$ can be obtained from $\gamma_j^{(0)}$
by, for $i=1$, the simultaneous permutation $u_2\leftrightarrow w_2$ and
$u_3\leftrightarrow w_3$; for $i=2$, $u_1\leftrightarrow w_1$ and
$u_3\leftrightarrow w_3$; and for $i=3$,  $u_1\leftrightarrow w_1$ and
$u_2\leftrightarrow w_2$.  This recipe produces the $\gamma_j^{(i)}$ with
a different indexing than in earlier work, but the value of $I_0$ is
not affected thereby.\\

There are problems with the numerical evaluation of $I_0$
 when the parameters $w_i$ and $u_i$ exactly or approximately
satisfy certain relationships;
these situations and methods for the avoidance of numerical 
instability have been
discussed elsewhere \cite{PRA,HFS}.

\end{document}